# Distribution of ranks of β-decay half-lives

Juan Miguel Campanario
Juan.campanario@uah.es

Departamento de Física
Universidad de Alcalá
28871 Alcalá de Henares
Madrid, Spain

**Abstract**

I studied the distribution of ranks of values of 2949 β-decay half-lives according to an empirical beta law with two exponents. β-decay half-life ranks showed good fit to a beta function with two exponents.

**Introduction**

The distribution of absolute values of variables in ranks has just started to receive attention from researchers. Martinez-Mekler et al studied the universal behavior of the manner in which elements of a system are distributed according to their rank with respect to a given property (1). The law they used is valid for the full range of values, regardless of whether or not a power law has previously been suggested. They proposed a two-exponent equation:

$$V(r) = K \frac{(N+1-r)^b}{r^a}$$

$V$ is the value of the property, $N$ is the number of cases and $r$ is the rank corresponding to the value $V$. In the above equation, $K$, $b$ and $a$ are the parameters that must be obtained.

Mansilla, Köppen, Cocho and Miramontes used the above empirical law to elucidate the rank-order behavior of journal impact factors (2). They found an extremely good fit that outperformed the earlier rank-order model. In previous research I have used this equation to study the rank-order behavior of changes in journal impact factors (3) and the rank of articles in and citations to journals (4). Martinez-Mekler et al (1) showed that the same law applies to other social and natural data.

Much attention has been paid to research on β-decay. However, it seems that less attention has been paid to the relationship between absolute values and ranks. Here I extend this previous work to explore the use of the equation suggested by Mansilla, Köppen, Cocho and Miramontes (2) to explain the rank-order distribution of β-decay half-lives. As in previous studies, the goal was to model the distribution of the absolute values (in this case β-decay half-lives) versus ranks.

**Method**

I used the experimental values of β-decay half-lives listed in the NUBASE database of nuclear and decay properties (5). I selected all values of β- and β+ decays, and converted half-lives to seconds. As in previous studies, the plot of ranks suggested that they did not fit a typical power law because of the downward bend in the tail on the right-hand side of the curve (Figure 1). Instead, I used the function suggested by Mansilla et al. (2).

Values given in the NUBASE table as approximations (e.g., ~456) were used as precise, unique data points (e.g, 456). Values given as "more than" or "less than" a certain cutoff value were also used as univocal values instead of approximate ranges (e.g., 230 rather than >230)

I transformed the equations using logarithms to yield:

$\log(HL) = \log(K) + b \log(N+1-r) - a \log(r)$

To obtain ranks of the half-lives, all records were listed from highest to lowest using Excel pivot tables to detect and remove duplicates (see, for example Campanario, 2009). I obtained N=2949 different ranks. Next, rank order was assigned from 1 to N. For example, the first values of β-decay half-lives in seconds, *HL (s)*, and rank, *r*, are

| Nuclide | Half-life (s) | Rank |
|---|---|---|
| $^{128}$Te | 6.943E+31 | 1 |
| $^{76}$Ge | 4.986E+28 | 2 |
| $^{130}$Te | 2.493E+28 | 3 |
| $^{82}$Se | 3.061E+27 | 4 |
| $^{48}$Ca | 1.673E+27 | 5 |

Next, I used the online service hosted at www.zunzun.com to obtain the parameters *K*, *b* and *a*. Once these parameters were obtained, the theoretical curve was plotted.

**Results and conclusions**

The parameters obtained for the linear equation are: $\log(K) = 19 \pm 3$; $b = 2.0 \pm 0.6$; $a = 7.5 \pm 0.6$ ($r^2 = 0.94$). Figure 1 plots the raw data and the results obtained with the model. As can be seen, the quality of the fit was very good.

The equation suggested by Mansilla et al. (2) has some advantages. For example, the parameter *a* is more influential for small values of *r*. When *r* is low, the law become nearly Lotkaian. However, as *r* increases, the influence of parameter *b* also does. This combination of influences can explain the decrease in half-life as rank increases.

In a recent paper, Egghe (6) presented a mathematical derivation of the rank-order distribution of journal impact factors. Egghe suggested that these distributions are valid for any type of impact factor (i.e., for any publication period and any citation period). According Egghe, the distributions may even be valid for average rank distribution of any sample. However, his analysis is based on the central limit theorem. According to Egghe's analysis, the distribution of the impact factors of journals in a given field is approximately normal, but according Waltman and van Eck (7) this is not the case.

Figure 2 shows the distribution of 99.3% of the records in intervals according the value of log half-life [log(HL)]. I have used log(HL) instead of HL because of the wide range of values for the variable HL. To simplify data presentation, I have excluded extreme values of log(HL). As can be seen, the values of β-decay half-lives are not distributed normally.

**Acknowledgments**

I thank K. Shashok for improving the use of English in the manuscript.

Figure legends

Figure 1 legend:
Semi-log rank order distribution of β-decay values.

Figure 2 legend:
Distribution of 99.3% of records in intervals according the value of log(HL). Extreme values of log(HL) have been excluded.

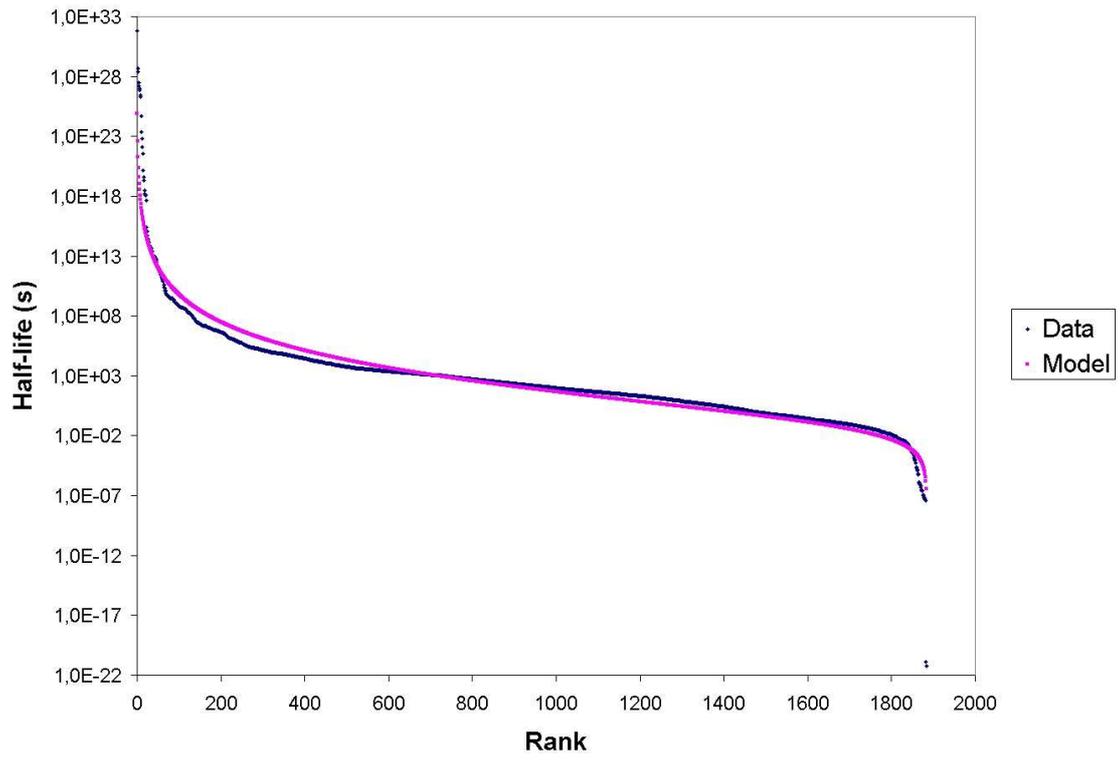

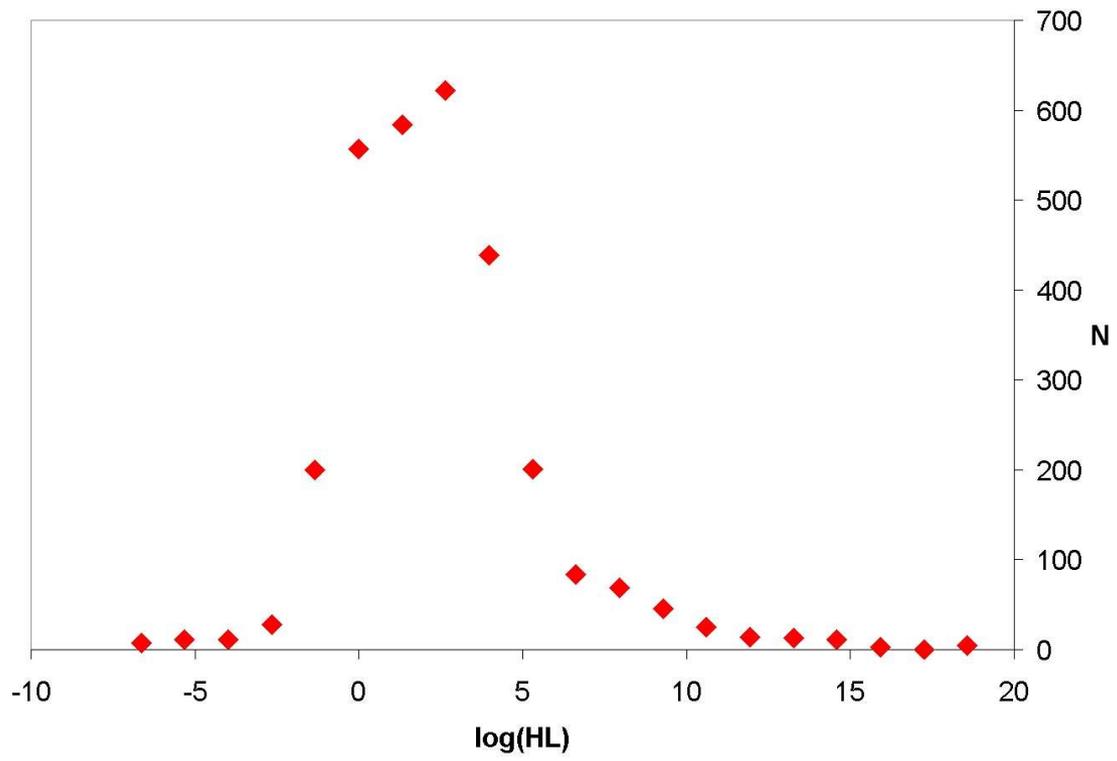